\documentclass[twocolumn,showpacs,preprintnumbers,amsmath,amssymb]{revtex4}

\usepackage{graphicx}
\usepackage{dcolumn}
\usepackage{bm}
\usepackage[usenames]{color}
\begin{document}

\newcommand{\x}[1]{{\bf #1}}

\title{Entanglement and Fidelity for Holonomic Quantum Gates}

\author{Paolo Solinas,$^{1,4}$ Maura Sassetti,$^{1, 3}$ Piero Truini,$^{1}$ Nino Zangh\`{\i}$^{1,2}$}
\affiliation{
$^1$ Dipartimento di Fisica, Universit\`a di Genova, \\
Via Dodecaneso 33, 16146 Genova, Italy \\
$^2$ Istituto Nazionale di Fisica Nucleare (Sezione di Genova), \\
$^3$ CNR-INFM Lamia, Genova \\
$^4$ Department of Applied Physics/COMP, Helsinki University of Technology P. O. Box 5100,
FI-02015 TKK, Finland \\
}

\date{\today}

\begin{abstract}

  We study entanglement and fidelity of a two-qubit system when a
  noisy holonomic, non-Abelian, transformation is applied to one of
  them. The source of noise we investigate is of two types: one due to
  a stochastic error representing an imprecise control of the fields
  driving the evolution; the other due to an interaction between the
  two near qubits.  The peculiar level structure underlying the
  holonomic operator leads us to introduce the reduced logical
  entanglement which is the fraction of entanglement in the logical
  space.  The comparison between entanglement and fidelity shows how
  they are differently affected by the noise and that, in general, the
  first is more robust than the latter.  We find a range of physical
  parameters for which both fidelity and reduced logical entanglement
  are well preserved.

\end{abstract}


\maketitle

\section{Introduction}

Entanglement is one of the most striking features of quantum
mechanics.  Its meaning and implications have been deeply discussed
since the early days of the theory but, in the last twenty years, its
study has found a new impulse due to its application to {q}uantum
{i}nformation \cite{rev_mod_phys}.  Entanglement has a crucial role
in quantum teleportation and quantum cryptography and it is believed
to be essential in quantum algorithms---the system must pass through a
state of maximum entanglement during the computation.  Since quantum
systems are very sensitive to perturbations and to the influence of
external noise, many efforts have been made in order to understand
when this precious resource may be lost and how it may be preserved
\cite{duan,carvalho,cai}.

With the aim of a robust and error free manipulation of quantum
information, geometric quantum computation has been proposed
about ten years ago \cite{jones,falci}.  According to this
  approach, the quantum information is processed by means of
operators depending only on global and geometric features of the
system's evolution. This feature has been shown to be powerful in
  creating and manipulating quantum information with high state
  fidelity \cite{high_fid}.  
In its non-Abelian version, called holonomic quantum computation,
various implementations have been proposed for systems such as trapped
ions \cite{unanyan,DCZ}, Josephson junction \cite{JJ}, Bose-Einstein
condensate \cite{BE}, neutral atoms \cite{neutral_atoms}, quantum dots
controlled by lasers \cite{solinas}. The main advantage of this approach {is} robustness of the single
qubit logical gate against both parametric and environmental errors.
Parametric errors are due to imprecise control of the {fields}
driving the system evolution; they have been shown to cancel out if
the control fields fluctuate fast enough
\cite{carollo1,dechiara,carollo2,kuvshinov,robustness_solinas}.  Also
in the case of environmental errors, the holonomic operators show an
intrinsic robustness \cite{parodi1,parodi2}.

In the present paper we {extend} the
  holonomic approach to a noisy two-qubit system in order to
  understand if and how the entanglement is preserved in holonomic
  quantum computation.  The question is not trivial even when, as in
the case of the present paper, only one of the two qubits undergoes a
holonomic transformation. In fact, it is typical of holonomic systems
{that the {\it logical space}, i.e., the space of degenerate states
  which go through a non-Abelian phase change, is embedded in a larger
  space \cite{wilczek-zee}. In our case the logical space is is
  embedded in a four dimensional space, {\it the system's Hilbert
    space}, as described in the next section. Since the evolution is
  not confined to the logical space, the entanglement cannot be
  estimated directly in terms of the von-Neumann entropy of the
  subsystems \cite{wootters}, rather one should first project the
  evolved quantum state onto the logical space. This projection leads
  to a non-unitary evolution in the logical space, that does not, in
  general, preserve the entanglement. We also calculate the fidelity
  and compare it with the entanglement, when the system is affected by
  noise.}

{We consider two types of noise: a parametric error that goes off
  whenever the driving fields are turned off, and a ``coupling error''
  due to undesired interactions between the two qubits.  We show that
  the holonomic operators are robust under parametric error. Moreover,
  we show that entanglement and fidelity are differently affected by
  these noises and that entanglement is more robust than fidelity. As
  one may expect, entanglement is influenced by the coupling between
  the qubits, however, we are able to single out a regime in which
  both entanglement and fidelity are preserved.}

The paper is organized as follows.  
{In Sec.~\ref{sec:HolEv} we give a
brief review of the holonomic approach to quantum computation; in
Sec.~\ref{sec:logical_space_ent} we define the {reduced logical
  entanglement} and we calculate  it 
in the presence of a parametric noise.  In
Sec.~\ref{sec:evolution_with_noise} we evaluate
the fidelity.  In Sec.~\ref{sec:coupled_qubs} we analyze both entanglement and fidelity
in the presence of an undesired coupling between the two
qubits.  {In Sec. \ref{sec:conclusions} we conclude}.}

\section{Holonomic Computation}
\label{sec:HolEv}
We briefly review the main features of {holonomic quantum
  computation {for a system (hereafter called system A) described by a four-dimensional Hilbert space $\mathcal{H}_A$, with a  level structure of  three
excited degenerate states $|i\rangle$ ($i=0,1,a$) at energy $\epsilon$,
and  ground state $|G\rangle$, set for convenience at energy 0}.  

The system is
driven by time-dependent  laser fields, with AC frequency in resonance with $\epsilon$, {inducing} transitions between
ground and excited states.  In the rotating frame representation the Hamiltonian governing the
evolution of the system is ($\hbar=1$) \cite{DCZ}
\begin{equation}
  H_0(t) = \sum_{i=0,1,a}\left(\Omega_i(t) |i\rangle \langle G| + h.c.\right)\,,
\label{eq:HQC_ham}
\end{equation}
where $\Omega_i(t)$ are the time-dependent Rabi frequencies of the laser fields. 
The Hamiltonian $ H_0(t)$  has four eigenstates: two {\it
  bright states} 
\begin{equation}
\left \{\begin{array}{lll}
  |B_{1} (t)\rangle &=& \frac{1}{\displaystyle\sqrt{2} \Omega(t) }
(\Omega(t) |G\rangle + \sum_i \Omega_i (t) |i \rangle) \\
	|B_{2} (t) \rangle &=& \frac{1}{\displaystyle\sqrt{2} \Omega(t)} 
(-\Omega(t) |G\rangle + \sum_i \Omega_i (t) |i \rangle) 
\end{array}
\right. \,, 
\label{eq:bright_states}
\end{equation}
where
\begin{equation}
\Omega(t)= \sqrt{\sum_{i=0,1,a} |\Omega_i(t)|^2} \; ,\label{energiepm}
\end{equation}
and two {\it dark states}
\begin{widetext}
\begin{equation}
\left\{\begin{array}{lll}
|D_1(t)\rangle &=& \frac{\displaystyle\Omega_a(t) (\Omega_1(t) |1 \rangle + \Omega_0 (t)|0 \rangle) -
 (\Omega(t)^2 - |\Omega_a(t)|^2) |a \rangle}{\displaystyle (\Omega (t)\sqrt{|\Omega_1(t)|^2+ |\Omega_0 (t)|^2})}
 \\
 |D_2(t) \rangle &=& \frac{\displaystyle \Omega_0(t) |1 \rangle - \Omega_1(t) |0 \rangle}
{\displaystyle \sqrt{|\Omega_1(t)|^2 + |\Omega_0 (t)|^2}}
\end{array}
\right. \,.
\label{eq:dark_states}
\end{equation}
\end{widetext}
The bright states have energy $E_\pm(t) = \pm \Omega(t)$ (the positive
value is associated with $|B_{1} \rangle$), while the dark states have
zero energy. {The time evolution  operator associated with (\ref{eq:HQC_ham}) is}
\begin{equation}
	U_0(t) ={\mathcal T} e^{-i\int_0^{t} d\tau {H_0}(\tau)}\;,
	\label{eq:time-dep}
\end{equation}
where ${\mathcal T}$ is the time-ordering operator.

{The modulation of the phase and the intensity of the laser fields
drives adiabatically the Hamiltonian along a loop in the
space of the Rabi frequencies---the parameter space---with $H_0(0)=H_0(T)$, where $T$ is the final adiabatic time. We shall assume that the Rabi frequencies have the following
time-dependence}
\begin{equation}
\left\{\begin{array}{lll}
 \Omega_0 (t)&=& {\Omega} \sin\theta(t) \cos\phi (t)\\ 
 \Omega_1 (t)&=& {\Omega} \sin\theta (t) \sin\phi (t) \\ 
 \Omega_a (t) &=& {\Omega} \cos\theta (t)
 \label{eq:param_loop}
\end{array}  \right. \,,
\end{equation}
{where  ${\Omega}$ is a constant. This means that we assume that the parameter space is the surface of a sphere of radius $\Omega$ and that  the adiabatic loop represented by (\ref {eq:param_loop}) is a closed curve on it.  Note
that the energies $E_\pm(t)$ of the bright states do dot depend on time and have constant value $\Omega$.} We shall
assume that $t=0$ and $t=T$ correspond to the north pole on the
sphere. {Thus, from (\ref{eq:bright_states}),  (\ref{eq:dark_states}) and (\ref{eq:param_loop}),}
\begin{eqnarray}
| D_1(0)\rangle&=&| D_1(T)\rangle=| 0\rangle\,,\label{cond1}\\ 
|D_2(0)\rangle&=&| D_2(T)\rangle=| 1\rangle\,, \label{cond2}\\ 
| B_1(0)\rangle&=&|B_1(T)\rangle=\frac{1}{\sqrt{2}}(| a\rangle+| G\rangle)\,,\label{cond3} \\
| B_2(0)\rangle&=&|
B_2(T)\rangle=\frac{1}{\sqrt{2}}(| a\rangle-| G\rangle)\,.
\label{condcontorno}
\end{eqnarray}

\begin{figure*}
	\centering
		\includegraphics[width=0.40\textwidth]{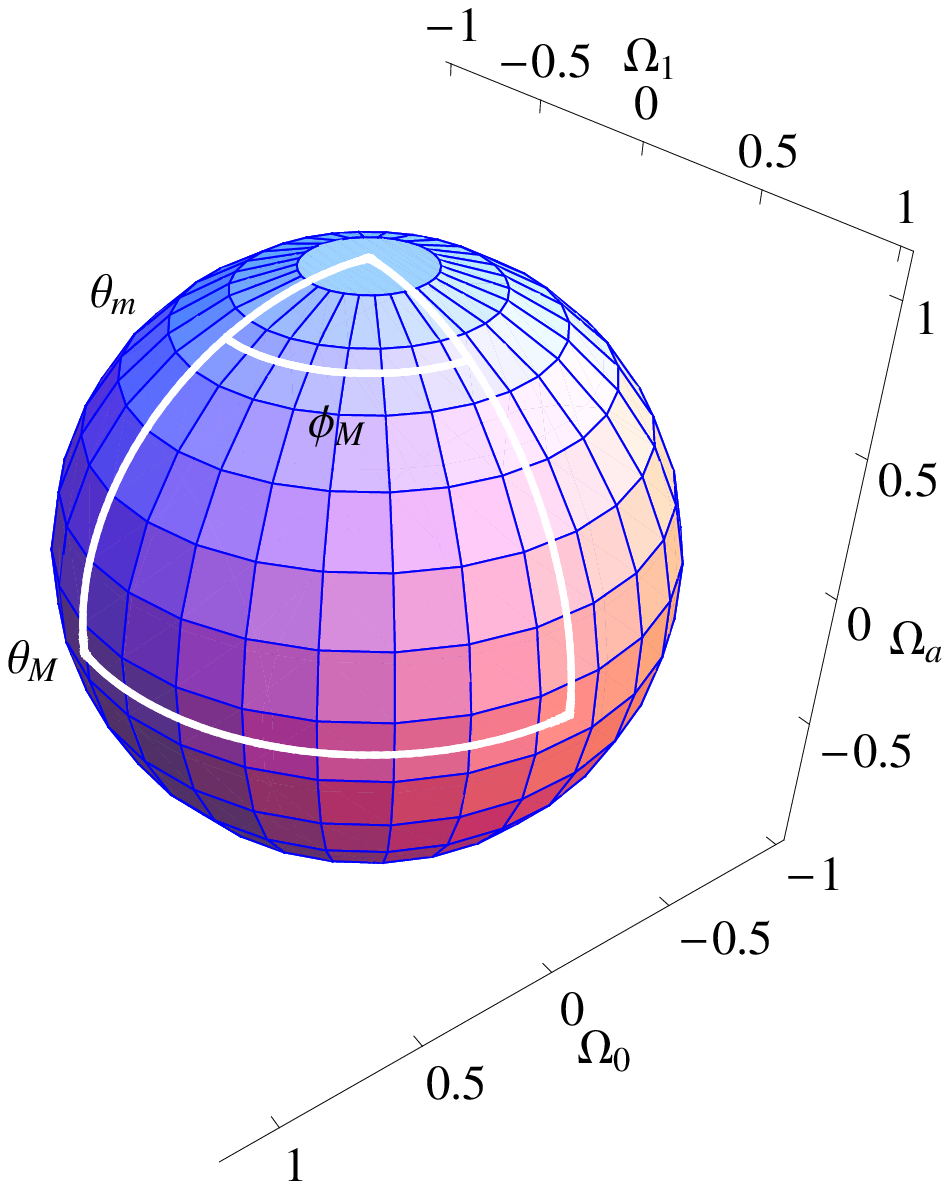}
		\includegraphics[width=0.40\textwidth]{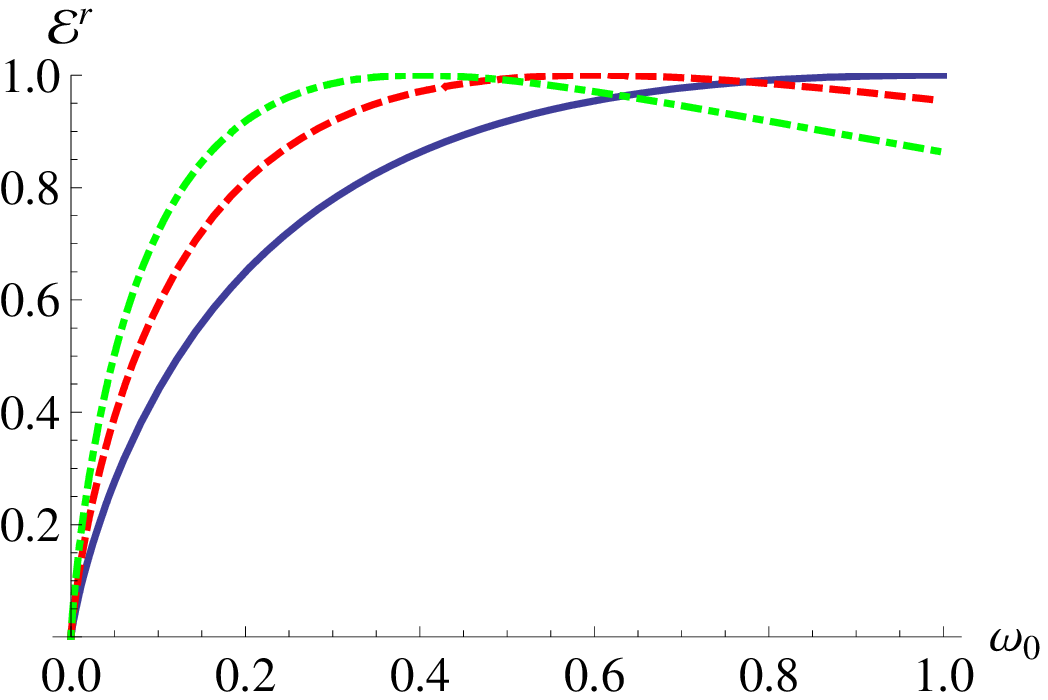}
                \caption{\label{fig:thetaphiloop}(color online) Left: Loop on the
                  sphere of radius $\Omega$ in the parameter space
                  which produces the desired logical operator with
                  $\eta=\pi /2$.  In the calculation of the integral
                  (\ref{eq:solid_angle}) the loop begins in
                  $(\theta_m,0)$ (with $\theta_m$ near to the north
                  pole), the evolution is along a meridian up to
                  $(\theta_M,0)$, then along a parallel up to
                  $(\theta_M,\phi_M)$, along a meridian again up to
                  $(\theta_m,\phi_M)$ and finally along a parallel
                  back to $(\theta_m,0)$.  We recover the desired
                  loop, that starts and ends at the north pole, in the
                  limit $\theta_m \rightarrow 0$.  The angle variables
                  are supposed to depend linearly on time.  Right:
                  Entanglement as a function of $\omega_0$, for
                  $\alpha=0$, with $\omega_1=0.4$ (dot-dashed), $0.6$
                  (dashed), $1$ (solid line).}
\end{figure*}

Under the adiabatic condition $\Omega T\gg 1$, the adiabatic theorem guarantees that any superposition of the dark states at $t=0$ will end up in another superposition of dark states at $t=T$, and that this transformation is realized by the unitary holonomic operator} \cite{DCZ}
\begin{equation}
	U_0(T)={\mathcal T} e^{-\int_0^{T} dt {A}(t)}\;,
	\label{eq:time-dep-HolOp}
\end{equation}
{obtained from (\ref{eq:time-dep}) in the adiabatic limit ${\Omega} T
\gg 1$}; ${A}(t)$ in (\ref{eq:time-dep-HolOp}) is the {\it connection} operator with 
{matrix elements} \begin {equation}
A_{ij}(t)= \langle D_i(t) | \frac{d}{dt} | D_j(t)\rangle\,,\quad{} i,j=1,2\,.
\label{aij}
\end{equation}
{Equation (\ref{eq:time-dep-HolOp}) is the core of the holonomic approach to quantum computation: in view of (\ref{cond1}), (\ref{cond2}) and (\ref{eq:time-dep-HolOp}), the states $|0\rangle$ and $|1\rangle$ can be regarded as {\it logical states} spanning the two-dimensional {\it logical space} $\mathcal{L}_A \subset \mathcal{H}_A$ (the ``A-qubit'') on which $U_0(T)$  acts as  {\it logical operator}.} 

{The main feature of the holonomic approach consists in the fact that $U_0(T)$
depends {\it only} on the area
  spanned by the curve (\ref{eq:param_loop}). To see how this comes about, let us evaluate the RHS of (\ref{eq:time-dep-HolOp}): for the matrix (\ref{aij}) one finds}
\begin{equation}
   A(t)= \left(
				\begin{array}{cc} 
                                  0 & -\dot{\phi (t)} \cos \theta(t) 
\\	\dot{\phi(t)} \cos \theta(t) & 0 
				\end{array}
\right)
\label{eq:connection}
\end{equation}
whence,
\begin{equation}
   U_0(T)= \left(
     \begin{array}{cc} 
       \cos \eta & \sin \eta \\					
       -\sin \eta & \cos \eta  
				\end{array}
	\right)\,,
	\label{eq:Hol_operator}
\end{equation}
where
\begin{equation}
\eta=\int_0^T dt \dot{\phi(t)} \cos \theta(t)
\label{eq:solid_angle}
\end{equation}
is the solid angle spanned on the sphere during the evolution. {Therefore, if} one wishes to construct, for example, a
  logical NOT operation, one has just to drive the external fields in
  such a way that the {closed curve (\ref{eq:param_loop})  spans} a solid
  angle $\eta=\pi/2$. In fact, for this value of $\eta$ , the matrix
  (\ref{eq:Hol_operator}) does what a NOT should do, namely, it
  exchanges the logical qubits $|0\rangle$ and $|1\rangle $ {(modulo a sign)}.

In order to evaluate the integral  (\ref{eq:solid_angle}),
we proceed in the following way:
\begin{enumerate}
\item {We choose a curve as in Fig.~\ref{fig:thetaphiloop}
    consisting in a sequence of evolutions along meridians and
    parallels where $\theta_{M}$, $\phi_{M}$ are the maximal angles
    spanned during the evolution. Therefore, the solid angle spanned
    by the curve is $\eta=\phi_{M}(1-\cos \theta_{M})$.}

\item {Since the north pole of the sphere is a singular point of the
    parametrization (\ref{eq:param_loop}),  we choose a loop along meridians and
    parallels which starts (and ends) at the angles $(\theta_{m},0)$,
    (Fig.~\ref{fig:thetaphiloop}).}  The value of the integral for the
  loop starting and ending at the north pole, is recovered in the
  limit $\theta_{m}\rightarrow0$ \cite{loopnote}.
\end{enumerate}

We shall now  include the effect of  parametric noise in the model.  
 In view of the adiabatic evolution, it is reasonable to assume that
the lasers are stable up to a relative error in the ratio between
noise and signal. {Thus} the effect of an imprecise control of the driving fields is easily
  modeled by replacing the Rabi frequencies $\Omega_i(t)$ in
  (\ref{eq:HQC_ham}) with $\widetilde{\Omega}_i (t)=
  \Omega_i(t) + \delta \Omega_i(t)$, {where $\delta \Omega_i(t)$ are random fluctuations
proportional to the intensities $\Omega_i(t)$ with $ \delta \Omega_i (t)\ll
\Omega_i (t)$.} {(From now on, for easiness of notation, we shall omit the explicit time dependence in the formulas and write, e.g., $ \delta \Omega_i $ instead of $ \delta \Omega_i (t)$.)}

To compute the perturbed geometric operator $\widetilde{U}(T)$ by means of
the perturbed connection $\widetilde{A}$, we  pass from the
Cartesian coordinates $\Omega_i +\delta \Omega_i$ to the spherical
ones,
\begin{eqnarray}
\widetilde{\Omega} &=& \Omega+\Delta \Omega\,,\nonumber\\
\widetilde{\theta}  &=& \theta +\Delta \theta\,,\nonumber\\
\widetilde{\phi}  &=& \phi+\Delta \phi \,.
\end{eqnarray}
From  (\ref{eq:param_loop}),  
\begin{eqnarray}
\widetilde{\Omega} &=&  \sqrt{\sum_i (\Omega_i+\delta \Omega_i)^2}\,,\nonumber\\
\widetilde{\theta}  &=& \arccos\left(\frac{\Omega_a+\delta \Omega_a}{ \Omega+\delta \Omega}\right)\,,\nonumber \\
\widetilde{\phi}  &=&  \arctan \left(\frac{\Omega_0 +\delta \Omega_0 }{\Omega_1+\delta \Omega_1}\right)\,.
\end{eqnarray}
By straightforward computation, one obtains the series expansion for
$\Delta \Omega$, $ \Delta \theta$ and $\Delta \phi$ in the small
parameters $\delta \Omega_i \ll \Omega_i $
\begin{eqnarray}
\Delta {\Omega} &=&  \delta \Omega + \delta^2 \Omega + O(\delta^3)\,,\nonumber\\
\Delta {\theta}  &=& \delta \theta + \delta^2 \theta+O(\delta^3)\,,\nonumber\\
\Delta {\phi}  &=&  \delta \phi + \delta^2 \phi+O(\delta^3) \;.
\end{eqnarray}
Here the ``$\delta^n$''-terms collect terms of order $n$ in $\delta \Omega_i$, e.g.,
\begin{eqnarray}
\delta \theta&=&\frac{\delta \Omega _0 \Omega _0 \Omega _a+\delta \Omega _1 \Omega _1
   \Omega _a+\delta \Omega _a \left(\Omega _a^2-\Omega
   ^2\right)}{\Omega ^2 \sqrt{\Omega _0^2+\Omega _1^2}} \,,
   \nonumber  \\
\delta \phi &=& \frac{\Omega_0 \delta \Omega_1-\Omega_1 \delta \Omega_0}{\Omega_0^2+\Omega_1^2} \,.
\label{eq:spherical_perturbations}
\end{eqnarray}
The perturbed solid angle is written as $$\widetilde{\eta}=\eta +\delta \eta +\delta^2 \eta +O(\delta^3)\,,$$
with  first order correction
\begin{equation}
\delta \eta = -\int_0^T dt (\dot{\phi} \delta \theta \sin \theta +  \delta \dot{\phi} \cos\theta ) \,.
\label{eq:angle_perturbation}
\end{equation} 

The perturbed dark states $ |\widetilde{D}_i\rangle$ are easily obtained
from (\ref{eq:dark_states}) by replacing $\Omega_i$ with
$\widetilde{\Omega}_i$.  Up to the second order in $\delta \Omega_i$ and
at the final time $t=T$, one finds
\begin{eqnarray}
  |\widetilde{D}_1\rangle&=&  \left(1-\frac{1}{2} \delta^2 \theta-\frac{1}{2} 
\delta^2 \phi\right) |D_1\rangle  \nonumber \\
  &+& (\delta \phi+\delta^2 \phi) |D_2\rangle - \frac{\delta \theta+\delta^2 \theta }{\sqrt{2}} 
\left(|B_1\rangle+|B_2\rangle\right)\,, \nonumber \\
  |\widetilde{D}_2\rangle &=&  -(\delta \phi+\delta^2 \phi)   |D_1\rangle + (1-\frac{\delta^2 \phi_1}{2})  
|D_2\rangle\,.
\end{eqnarray}
{Note that} $|\widetilde{D}_1\rangle$ {is a}
superposition of both unperturbed dark and bright states thus leading
to a population {\it leakage} from the unperturbed dark space to the
unperturbed bright space.

The perturbed final operator $\widetilde{U}(T)$, written in the {basis} 
$\{ |\widetilde{D}_1\rangle, |\widetilde{D}_2\rangle\}$, can be read
directly from equation (\ref{eq:Hol_operator}), {by replacing $\eta$ with} 
$\widetilde{\eta}= \eta +\delta \eta$. In particular, for $\eta=\pi/2$, up to second order, 
\begin{equation}
\widetilde{U}(T)=\left(
\begin{array}{ll}
 -\delta \eta-\delta^2 \eta &\quad 1-\frac{(\delta \eta)^2}{2} \\\\
 -1+\frac{(\delta \eta)^2}{2} &\quad -\delta \eta-\delta^2 \eta
\end{array}
\label{eq:tildeU}
\right) \,.
\end{equation}

\section{Entanglement in multilevel systems}
\label{sec:logical_space_ent}

{We shall now analyze a two-qubit system.  We begin by specifying the model that we shall investigate}:  \begin{enumerate}
\item {We consider the composite system formed by two replicas of the system considered in the previous section---one will be called system A and the other system B. The Hilbert space of the composite is then $\mathcal{H}_A\otimes \mathcal{H}_B$, where $\mathcal{H}_B$, the Hilbert space of system B, is a replica of  $\mathcal{H}_A$.} 

\item {The logical space of the composite is $\mathcal{L}_A\otimes \mathcal{L}_B$, where, as before, $\mathcal{L}_A$ is the logical space of system A, spanned by the logical states $|0\rangle$ and $|1\rangle$, and $\mathcal{L}_B$ is the replica of $\mathcal{L}_A$ contained in $\mathcal{H}_B$. Following the standard terminology, we shall call $\mathcal{L}_A$ the A-qubit and $\mathcal{L}_B$ the B-qubit.}

\item {We shall assume that {\it only} the A-qubit  undergoes information processing (for sake of concreteness, we shall assume that  the A-qubit performs a  logical NOT 
operation).}

\item {The initial state of the system is taken to be the maximally
entangled state} \begin{equation} \Psi = \frac{1}{\sqrt{2}} (|00\rangle
  + |11\rangle)\label{maxent}\,. \end{equation}

\end{enumerate}
{We note that the relevant simplifying assumption is 3, while 4 is rather standard for the kind of problem we wish to address here (see below). Note that assumption 3 allows two possibilities for representing the time evolution of the two-qubit system:} \begin{itemize}
\item[a)] {a unitary transformation of the form  $U\otimes I$, where $U$ governs the
dynamics of the system A (e.g., as given by (\ref{eq:tildeU})), and $I$ is the identity in  $\mathcal{H}_B$;} \item[ b)] {a unitary transformation which accounts for the interaction between the two qubits.} 
\end{itemize}
{Case a) will be addressed in this section and case b) in Sect. \ref{sec:coupled_qubs}.}

{Our aim is to study how the entanglement of the two qubits is
preserved by the dynamics and to do this we need a quantitative estimate of the entanglement. 

When a system is composed by two sub-systems A and B,
the entanglement of the composite system, in the case of pure states,
can be estimated in terms of the von Neumann entropy} 
\begin{equation}
 \mathcal{E}=-\mbox{Tr}(\rho_A \log_2 \rho_A)=-\mbox{Tr}(\rho_B \log_2 \rho_B)\,.\label{entropia}
\end{equation}
{where  $\rho_A$ and $\rho_B$ are the reduced density matrices 
 of the sub-systems \cite{wootters}.}
{However, since in our model the qubits are embedded in a larger space, formula (\ref{entropia}) cannot be directly applied.} 

{To arrive at a suitable estimator of entanglement, we  proceed as follows: Let  $P$ denote the projector on the logical subspace $\mathcal{L}_A$  of  $\mathcal{H}_A$  and  let $P\otimes P$ be the projector onto $\mathcal{L}_A\otimes \mathcal{L}_B$. Consider}
\begin{equation}\rho^r = P \otimes P 
  \rho P\otimes P \end{equation} 
{and} \begin{equation}
\rho_A^r=\mbox{Tr}_B \rho^r\,.
\label{iar}\end{equation}
{Then the quantity}
\begin{equation}\mathcal{E}^r=-\mbox{Tr} (\rho_A^r \log_2
  \rho_A^r)\,, \label{entropiar}
\end{equation}
{is analogous to (\ref{entropia}), and can be regarded as an estimator  of the fraction of entanglement in the logical
subspace (see also \cite{nihira-stroud}).  Hereafter we shall refer to 
to $\mathcal{E}^r$ as  the {\it reduced
  logical entanglement} or, when no ambiguity will arise, 
simply as the {\it entanglement}.}

{We wish now to obtain a convenient formula for
  $\mathcal{E}^r$ {(under the assumption a) specified above)}. Firstly, we evaluate \begin{equation} \rho_A^r=
    \mbox{Tr}_B \left( P \otimes P U\otimes I |\Psi\rangle\langle
      \Psi| U^{*}\otimes I P\otimes P \right) \label{eq:ar}
 \end{equation}
without  relying on any specific form of the unitary $U$. 
{To this end we note that a
generic} transformation  $U$ of the logical qubit basis can  be written as
\begin{eqnarray}
 U |0 \rangle &=& P U |0 \rangle + P^\perp U |0 \rangle  \nonumber \\
 U |1 \rangle &=&   P U |1 \rangle + P^\perp U |1 \rangle\;.
  \label{eq:noisy_ev}
\end{eqnarray}
Let  \begin{equation}
\sqrt{\omega_0} = \left\|PU |0\rangle \right\|\;, \;\; \sqrt{\omega_1} = \left\|PU |1\rangle \right\|
\end{equation}
and denote by $|\psi_0\rangle$, $|\psi_1\rangle$, $ |\psi_0^{\bot}
\rangle$, and $|\psi_1^{\bot}\rangle$ the vectors obtained by
normalization from $P U |0 \rangle $, $ P U |1 \rangle$, $P^{\bot} U
|0 \rangle $, and $ P^\perp U |1 \rangle$ respectively. Then
\begin{eqnarray}
 U |0 \rangle &=&  \sqrt{\omega_0} |\psi_0 \rangle +
   \sqrt{1-\omega_0} |\psi_0^{\bot} \rangle \nonumber \\
 U |1 \rangle &=&  \sqrt{\omega_1} |\psi_1 \rangle +\sqrt{1-\omega_1} |\psi_1^{\bot} \rangle\,.
  \label{eq:noisy_ev1}
\end{eqnarray}
{It is useful} to introduce the scalar product
\begin{equation}
\alpha =\langle \psi_0 |\psi_1 \rangle\label{alpha}\,.
\end{equation}
Note that the coefficients  $\sqrt{1-\omega_i}$ represent the leakage of the populations 
and that in general $\alpha$ is different from zero.
{Therefore}
\begin{eqnarray}
U\otimes I|\Psi\rangle	&=& \frac{1}{\sqrt{2}} \left[ 
\sqrt{\omega_0} |\psi_0  0\rangle +\sqrt{1-\omega_0} |\psi_0^{\bot} 0 \rangle \right. \nonumber \\
	&& \left. + \sqrt{\omega_1} |\psi_1 1\rangle+ \sqrt{1-\omega_1} |\psi_1^{\bot} 1\rangle\right] 
	\,,
\end{eqnarray}
{and}
\begin{equation}
PU\otimes I|\Psi\rangle  = \frac{1}{\sqrt{\omega_0+\omega_1}} [\sqrt{\omega_0} |\psi_0 0\rangle + \sqrt{\omega_1} 
|\psi_1 1\rangle] \,.
\end{equation}
Noting that $|\psi_1\rangle$ can be decomposed into its components
along and orthogonal to $|\psi_0\rangle$: $|\psi_1\rangle= \alpha
|\psi_0\rangle + \beta |\widetilde{\psi}_0 \rangle$ with $\langle \psi_0|\widetilde{\psi}_0\rangle=0$,
$\alpha$ given in (\ref{alpha}), 
and $\beta=\sqrt{1-\alpha^{2}}$, we find
\begin{equation}  PU\otimes I|\Psi\rangle  = 
	\frac{\sqrt{\omega_0} |\psi_0 0\rangle + 
	\alpha \sqrt{\omega_1} |\psi_0 1\rangle + \beta \sqrt{\omega_1} |\widetilde{\psi}_0 1\rangle}
{\sqrt{\omega_0+\omega_1}}\,.
	\label{eq:proj_state}
\end{equation}
{Thus, we may evaluate   (\ref{eq:ar}) in  the $\{|\psi_0 \rangle\,, |\widetilde{\psi}_0 \rangle\}$ basis,} 

\begin{equation}
 \rho_A^r=\frac{1}{\omega _0+\omega _1}
\left(
\begin{array}{ll}
\omega _1 \alpha^2+\omega_0 &\quad
   \alpha  \sqrt{1-\alpha^2} \omega_1 \\\\
 \alpha  \sqrt{1-\alpha^2} \omega_1 &\quad
   -\left(\alpha^2-1\right) \omega_1
\end{array}
\right)  \,,	
\end{equation}
{whence}
\begin{equation}
	\mathcal{E}^r=-(\lambda_- \log_2 \lambda_- + \lambda_+ \log_2 \lambda_+)\,,
	\label{eq:entanglement}
\end{equation}
where
\begin{equation}
 \lambda_{\pm}= \frac{\omega _0+\omega _1 \pm \sqrt{\omega _0^2+2 \left(2 \alpha
   ^2-1\right) \omega _1 \omega _0+\omega _1^2}}{2 \left(\omega _0+\omega _1\right)} 
   \label{lambdapm}
\end{equation}
{are the  eigenvalues of $ \rho_A^r$. }

In Fig.~\ref{fig:thetaphiloop} (right),  $\mathcal{E}^r$ is plotted
  as a function of $\omega_0$ for $\alpha=0$ and different values of
$\omega_1$. {It follows immediately from (\ref{eq:entanglement}) and (\ref{lambdapm}) that  $\mathcal{E}^r =1$ for  $\alpha=0$ and $\omega_0= \omega_1$ (this includes the trivial case of no leakage for which $\omega_0=\omega_1=1$, as well as  non-trivial cases  of population leakage with $\omega_0=\omega_1<1$).  For $\alpha \neq 0$, it is  always  $\mathcal{E}^r < 1$}.

\label{sec:ent_fid_cal}

We now {specialize to the case  $U=\widetilde{U}(T)$, with}  $\widetilde{U}(T)$ given by (\ref{eq:tildeU}).  Keeping
the contributions up to the second order in the perturbations $\delta
\Omega_ i$, we obtain
\begin{eqnarray}
P \widetilde{U}(T) | 0\rangle &=& \sqrt{\omega_0} |\psi_0 \rangle =
 \left(\delta \phi+\delta^2 \phi -\delta \eta-\delta^2 \eta \right) |0\rangle  \nonumber \\  
 &&+
 \left(\frac{\delta \phi^2}{2}-\delta \eta \delta \phi +\frac{(\delta \eta)^2}{2}-1 \right)|1\rangle  
\nonumber \\
P^\perp \widetilde{U}(T) | 0\rangle &=& \sqrt{1-\omega_0}|\psi_0^{\bot}\rangle = \frac{\delta \theta  
\delta \eta}{\sqrt{2}} \left(|a \rangle +|G\rangle \right) 
 \nonumber\\
  P \widetilde{U}(T) | 1\rangle &=&\sqrt{\omega_1}|\psi_1 \rangle \nonumber \\
  &=& \Big(
  1-\frac{\delta \theta^2}{2}-\frac{\delta \phi^2}{2}-\frac{(\delta \eta)^2}{2}+\delta \phi 
   \delta \eta  \Big) |0 \rangle  \nonumber \\
   &+&    \Big( \delta \phi+\delta^2 \phi -\delta \eta-\delta^2 \eta \Big) |1 \rangle \nonumber \\
P^\perp \widetilde{U}(T) | 1\rangle &=& \sqrt{1-\omega_1}|\psi_1^{\bot}\rangle  \nonumber\\
&=& -\frac{\delta \theta+\delta^2 \theta }{\sqrt{2}}  \left(|a \rangle +|G\rangle \right)
  		 \label{eq:psi_1_T}
\end{eqnarray}
and
\begin{eqnarray}
	\alpha &=& O(\delta^3) \nonumber \\
	\omega_0 &=& 1+O(\delta^4) \nonumber \\
	\omega_1 &=& 1-\frac{(\delta \theta)^2}{8} +O(\delta^3)\,.
 \label{eq:leakage}
\end{eqnarray}
{In the above equations $\delta \theta$ and $\delta\phi$ are evaluated at $t=T$.
From (\ref{eq:param_loop}) and (\ref{eq:spherical_perturbations}) it follows that $\delta\theta=0$ and that one can set $\delta\phi=0$. Thus  $\alpha=0$ and
$\omega_0=\omega_1=1$, whence, as above, $\mathcal{E}^r =1$.}

\section{Fidelity}
\label{sec:evolution_with_noise}
If the system is subject to noise sources, the fidelity
  $\mathcal{F}$ is used to quantify the ``distance'' between the
  perturbed final state $ \mathcal{U}(T) \Psi$, due to a perturbed evolution $\mathcal{U}$,
  and the final unperturbed state  $U_0
      (T)\otimes I\Psi$, with $U_0(T)$ given by
  (\ref{eq:Hol_operator}). For composite systems,   $\mathcal{F}$  provides an estimation of the
performance of logical operators that supplement the information
already provided by $\mathcal{E}^r$. The explicit formula for the fidelity  is
  \begin{equation}\label{fidelity1}
    \mathcal{F} = \left | \langle \mathcal{U}(T) \Psi | U_0
      (T)\otimes I\Psi\rangle\right |\,.
\end{equation}
{For the case we wish to consider first, $ \mathcal{U}(T) = \widetilde{U}(T)\otimes I $,  with  $\widetilde{U}(T)$ given by (\ref{eq:tildeU})}. 
{From (\ref{eq:psi_1_T}), by a straightforward computation, one
  obtains}
\begin{equation}
 \mathcal{F} = 1-\frac{(\delta \theta)^2}{4}-\frac{(\delta \phi)^2}{2}
 -\frac{(\delta \eta)^2}{2}+\delta \phi \delta \eta + O(\delta^3)\,.
 \end{equation}
 {Since, the terms $\delta \theta$ and $\delta \phi$ are evaluated
   at time $T$, their contribution is zero and thus}
\begin{equation}
 \mathcal{F} \approx  1 -\frac{(\delta \eta)^2}{2} \,.
 \label{eq:fidelity}
\end{equation}
{The fidelity depends on the error $\delta \eta$ relative to
  the area spanned during the perturbed evolution, it then appears to be
  less robust than entanglement.}

\label{sec:random_noise}

{We shall now estimate $\delta \eta$ for a simple model of 
  noise. We first integrate by parts the
  second term in Eq.~(\ref{eq:angle_perturbation}) neglecting the
  contribution at the endpoints}
\begin{equation}
 \delta \eta = -\int dt (\dot{\phi} \delta \theta + \dot{\theta} \delta \phi )\sin \theta \,.
\end{equation}
{We consider the loop in Fig.~\ref{fig:thetaphiloop}} passing near
the north pole and then take the limit for $\theta_m \rightarrow 0$,
in the notation of the figure caption.  We separate the contribution
of four different parts depending on the evolution along meridians and
parallels (cf. Fig.~\ref{fig:thetaphiloop}) and denote by $\int_i f(t)
dt$ the contribution along the $i$th path.  We have
\begin{eqnarray}
  \delta \eta &=& -\int_{1,3} dt \dot{\theta} \delta \phi  \sin \theta   
- \sin \theta_{M} \int_2 dt \dot{\phi} \delta \theta \nonumber \\
  && - \sin \theta_{m} \int_4 dt \dot{\phi} \delta \theta \,.
\end{eqnarray}
{Note} that in the limit of $\theta_m \rightarrow 0 $ the fourth integral gives no contribution.
{We now suppose that the perturbations $\delta \Omega_i$ fluctuate
  randomly over a time scale $\tau \ll T$ and go off whenever the
  driving fields are turned off, i.e. that they are of the
  form\begin{eqnarray}
    \delta \Omega_i(t) &=&   \Omega_i(t)  Z_i(t) = \nonumber \\
    &=& \Omega_i(t) \sum_{k=0}^{n-1} h \left (\frac{t- k \tau}{\tau}
    \right) Z^i_k \,,
 \label{eq:noise_expression}
\end{eqnarray}
where $h ( x )$ is the ``box'' function, which is equal to $1$ in the
interval $0 \leq x < 1$ and zero elsewhere, $n=T/\tau$, and $ Z^i_k$
are independent Gaussian} random variables, with zero average and
variance $\sigma^2$.

Using (\ref{eq:param_loop}) and (\ref{eq:spherical_perturbations}) we
can explicitly write the contributions for the separate evolutions.
Along the first meridian we have no contribution since $\phi = 0 $.
The only non-zero contributions are along the second and third paths;
we suppose that $\theta$ and $\phi$ depend linearly on time (see also
\cite{loopnote}) and write $\dot{\theta}=v_\theta$ and
$\dot{\phi}=-v_\phi$. We thus obtain
\begin{align}
 \delta \eta &=  v_\phi \cos \theta _{M} \sin ^2\theta _{M }   
\int_2 dt  \left(Z_0 \cos ^2\phi + Z_1\sin ^2\phi  -Z_a\right)+ \nonumber \\
 &- \frac{v_\theta}{2} \sin 2 \phi_M  \int_3 dt \left(Z_0-Z_1\right) \sin \theta\,.   
\label{eq:stoc_error}
\end{align}
Inserting (\ref{eq:noise_expression}) in (\ref{eq:stoc_error}) one can perform the integration and the
expansion at first order in $\tau/T=1/n \ll 1$
\begin{align}
 \delta \eta &= 4 \phi_M  \cos \theta _M \sin ^2 \theta _M\times\nonumber \\
 & \times\sum_{k=0}^{n-1} \Big[ \frac{Z^{0}_k \cos ^2( k \tau  v_\phi)}{n}
-\frac{ Z^{1}_k\sin ^2  (k \tau  v_\phi)}{n}-\frac{Z^{a}_k}{n}\Big ]+  \nonumber \\
 & -2  \left(\theta_M-\theta _m\right) \sin \left(2 \phi _M\right)
\sum_{k=0}^{n-1} \frac{\sin \left(k \tau v_{\theta }\right)
   \left(Z_{k}^0-Z_{k}^1\right)}{n}\,. 
\end{align}
{Using the independence of the random variables $ Z_{k}^{i}$, from
  the {\it central limit theorem} it follows that for $n\gg
  1$} 
\begin{equation} (\delta \eta)^2 \approx \frac{\sigma^2}{n} =
  \frac{ \sigma^2 \tau}{T}\,,
\label{erroresueta}
\end{equation}
which expresses the {\it cancellation effect} already discussed in
\cite{robustness_solinas} for large $n$.  Thus the fidelity
(\ref{eq:fidelity}) becomes
\begin{equation}
\mathcal{F} \approx 1- \frac{\sigma ^2}{2}\frac{\tau}{T}\,.
\end{equation}

\begin{figure}
	\centering
        \includegraphics[width=0.40\textwidth]{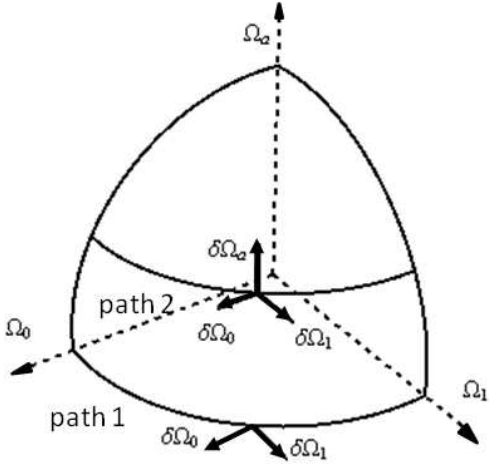}
	\caption{Evolution along the equator (path $1$ with
          $\theta_M=\pi/2$) and along a generic parallel (path $2$).
          In the first case, only the perturbation $\delta \Omega_0$
          and $\delta \Omega_1$ for the laser turned on are present.}
	\label{fig:pert_loop}
\end{figure}

The above calculation is valid for any loop in the parameter space
which moves along a parallel and meridian and spans a $\pi/2$ solid
angle.  Among these loops, there is one which is particularly
interesting.  If $\theta_M=\pi/2$ and $\phi_M=\pi/2$, we are moving on
the $\Omega_a-\Omega_0$ plane, along the equator and on the
$\Omega_a-\Omega_1$ plane.  As can be seen from (\ref{eq:stoc_error}),
with this loop the contributions along the second and third paths are
zero and $\delta \eta=0$ independently of the characteristics of the
noise (variance and correlation time).  In other words, this loop
makes the system completely robust against this particular
perturbation.  This fact has a simple geometrical and physical
interpretation.  We recall that the loop describes the way we turn on
and off the lasers.  The above loop is the one in which we have always
one laser completely turned off while modulating the intensities of
the other two; for example, in Fig.~\ref{fig:pert_loop} is shown the
evolution along the equator (path $1$) when $\Omega_a$ laser turned
off and we are modulating the $\Omega_0$ and $\Omega_1$ intensities.
Along this path, $\delta \Omega_a=0$ while we have perturbation of the
other two lasers.  However, the perturbations $\delta \Omega_0$ and
$\delta \Omega_1$ produce only radial and $\delta \theta$ perturbation
which do not affect the solid angle along this path.  In other words,
choosing this particular loop, along the single paths, we eliminate
the part of perturbation which can modify the solid angle spanned and
as result the angle is spanned without error.

\section{Coupled Qubits}
\label{sec:coupled_qubs}

In a more realistic situation, the two qubits can also interact.  This
interaction allows to manipulate the two systems as a whole and it is the
basis for constructing two qubit gates.  Ideally, one can control the
coupling strength and turn it on and off depending on the logical
gate.  However, if the interaction cannot be perfectly turned off, its
presence results in a new source of noise. Here, we choose
to describe it with a simple model with
\begin{equation}
H_I = \chi |11\rangle \langle 11| \,.
\label{interaction}
\end{equation}
As before,  only the A qubit undergoes an information process.  
 Due to the
specific form of the coupling (\ref{interaction}) only the state
$|1\rangle$ of qubit B will feel the additional interaction.  It is
then convenient to analyze the evolution in the dark-bright
basis for the qubit A and in the state $|1\rangle$ for the quit B:
$\{|D_1 1\rangle, |D_2 1\rangle, |B_1 1\rangle, |B_2 1\rangle \}$
(neglecting, for the moment, additional errors induced by the imprecise
control). If
$\Omega \gg \chi$, and the system starts in a superposition of dark
states, the transition to the bright states are negligible and the
evolution stay in the dark space $\{|D_1 1\rangle, |D_2 1\rangle
\}$.  The Hamiltonian $H_I$ in this basis is
\begin{equation}
H_I=\chi \left(
\begin{array}{ll}
 \cos ^2\theta \sin ^2\phi  &\quad \cos \theta  \cos \phi \sin \phi  \\\\
 \cos \theta \cos \phi  \sin \phi  &\quad \cos ^2 \phi 
\end{array}
\right).
\label{eq:H_IDarkBright}
\end{equation}

Thus, one sees that $H_I$ breaks the energy degeneracy $E_1=E_2=0$
of the two dark states $|D_1\rangle$ and $ |D_2\rangle$, relative to the unperturbed Hamiltonian 
(\ref{eq:HQC_ham}). The new eigenvalues are easily evaluated by
diagonalization in the presence of $H_I$, 
\begin{equation}
\bar E_1=0\,,\quad \bar E_2 = \chi (\cos^2\phi +\cos^2\theta \sin^2\phi)\,.
\label{newE}
\end{equation}
The corresponding eigenvectors can be written as a superposition of the unperturbed dark states
\begin{eqnarray}
      |{\bar{D}}_1(t)\rangle &=& \alpha(t)|D_1(t)\rangle -\beta(t)|D_2(t)\rangle\nonumber\\ 
      |{\bar{D}}_2(t) \rangle &=& \beta(t)|D_1(t)\rangle + \alpha(t)|D_2(t)\rangle\,,  
\label{eq:dark_states_H_I}
\end{eqnarray}
with 
\begin{eqnarray}
\alpha(t) &=& \frac{\displaystyle{\cos\phi(t)}}
{\displaystyle{\sqrt{\cos^2 \phi(t)+\cos^2 \theta(t) \sin^2\phi(t)}}}\nonumber\\
\beta(t) &=& \frac{\displaystyle{\cos\theta(t) \sin\phi(t)}}
{\displaystyle{\sqrt{\cos^2 \phi(t)+\cos^2 \theta(t) \sin^2\phi(t)}}}\,.
\label{alphaebeta}
\end{eqnarray} 
These coefficients satisfy boundaries conditions, related to the
closed loop in Fig.~(\ref{fig:thetaphiloop}) with $\theta(0)=\theta(T)=\phi(0)=\phi(T)=0$
\begin{equation}
\alpha(0)=\alpha(T)=1\,,\quad \beta(0)=\beta(T)=0\,,
\label{boundaries}
\end{equation}
and normalization
\begin{equation}
\alpha^2(t)+\beta^2(t)=1.
\label{norm1}
\end{equation}
Note that the energy shift (\ref{newE}), induced by the perturbation, may produce important
modifications during the time evolution since non-Abelian effects are based on the
assumption of degeneracy. If $\chi T \gg 1 $ the perturbed dark states are
separated during the evolution and no holonomic operator
(\ref{eq:time-dep-HolOp}) can be produced. For this reason we shall focus, in the following, 
on the more relevant  case $\chi T \leq 1$, which preserve the logical space.  

The evolution operator is now
\begin{equation}
U_\chi(T)=\mathcal{T} e^{-i\int_0^T\left[H_{I}(t)-i\bar{A}(t)\right]dt}\,.
\label{uchi}
\end{equation}
{(Note that the first term in the exponential, which represents the dynamic contribution of the 
perturbed dark states,  is absent in  (\ref{eq:time-dep-HolOp})  since there the dark states are degenerate)}. Using the new basis
$\{|{\bar{D}(t)}_1 1\rangle, |{\bar{D}(t)}_2
1\rangle\}$, $$\bar{A}_{ij} (t)= \langle
{\bar{D}}_i (t)1| \frac{d}{dt} |
{\bar{D}}_j (t) 1\rangle\quad i,j=1,2\,,$$ and thus
\begin{equation}
\bar{A(t)}= \left(
  \begin{array}{ll}
    \alpha \dot{\alpha}+\beta \dot{\beta}  & 
    \quad \beta \dot{\alpha}- \alpha \dot{\beta} 
    -\cos \theta \dot{\phi}  \\
    \alpha \dot{\beta}-\beta \dot{\alpha}+\cos \theta \dot{\phi} 
&\quad \alpha \dot{\alpha}+\beta \dot{\beta}
\end{array}
\right)\,.
\label{eq:connection_coupled}
\end{equation}

{The integral $\int_0^T \bar{A}(t)dt$ in} (\ref{uchi}) can be evaluated
using (\ref{alphaebeta}), (\ref{boundaries})
and (\ref{norm1}). The integrals of the diagonal terms in
(\ref{eq:connection_coupled}) are always zero because they represent
the derivative of the norm of the dark states, which is time
independent.  For the off-diagonal part, it can be easily {shown} that $\alpha
\dot{\beta} -\beta \dot{\alpha} =\dot{\alpha}/(1-\alpha^2)^{1/2}$ gives no contribution
once it is integrated along the closed loop.  Thus
\begin{equation}
\int_0^T \bar{A}(t)dt= \left(
\begin{array}{ll}
 0&\quad -\eta  \\
 \eta &\quad\;0
\end{array}
\right)\,.\label{aa}
\end{equation}

{The integral $\int_0^T H_{I}(t) dt$ in (\ref{uchi}) can be evaluated  by performing the integration} along the loop shown in Fig.~\ref{fig:thetaphiloop} with constant velocities.  One {obtains} 
\begin{equation}
\int_0^T H_{I}(t) dt= \int_0^T \bar E_2 dt = \chi T \gamma(\theta_M, \phi_M)\label{hh}
\end{equation}
with
\begin{eqnarray}
 \gamma(\theta_M, \phi_M)&=&
       \frac{1}{3} + \frac{\sin 2 \theta _M \sin ^2\phi_M
       +\left(3+\cos 2 \phi_M\right) \theta_M}{12 \theta_M}\nonumber\\
   &+& \frac{\sin 2 \phi_M \sin ^2\theta_M +\left(3+\cos 2\theta_M \right) 
   \phi_M}{12 \phi_M}\,.
\label{gamma}
\end{eqnarray}
Here, $\theta _M$ and $\phi _M$ satisfy $\phi_{M}(1-\cos \theta_{M})=\eta$,
with $\eta$ the solid angle spanned on the sphere during the evolution
given in (\ref{eq:solid_angle}).

{From (\ref{aa}) and (\ref{hh}), one may evaluate the RHS of (\ref{uchi}) and obtain, in the logical basis $\{|0\rangle, |1\rangle\}$, the matrix}
\begin{equation}
U_\chi(T)=\frac{1}{\mu}
\left(
\begin{array}{ll}
 e^{i \frac{\chi T \gamma}{2} }  K_-&\quad 2 e^{i \frac{\chi T \gamma}{2}}
   \eta  \sin \frac{\mu }{2} \\\\
 -2 e^{i \frac{\chi T \gamma}{2}} \eta  \sin \frac{\mu }{2} &\quad
   e^{i \frac{\chi T \gamma}{2}} K_+
\end{array}
\right),
\label{eq:U_chi}
\end{equation}
where 
\begin{equation}
K_{\pm}=\left(\mu  \cos \frac{\mu }{2}\pm i \chi T \gamma
   \sin \frac{\mu }{2}\right),\quad \mu= \sqrt{4 \eta ^2+(\chi T \gamma) ^2}\,.
\end{equation}

\begin{figure*}
	\centering
        \includegraphics[width=0.40\textwidth]{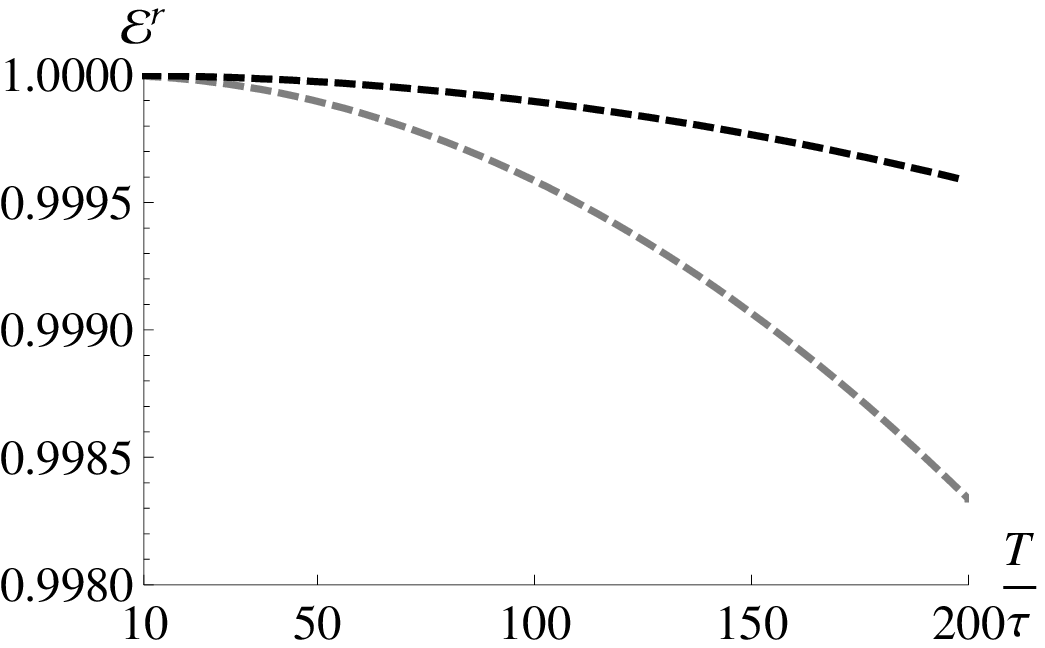}
		\includegraphics[width=0.40\textwidth]{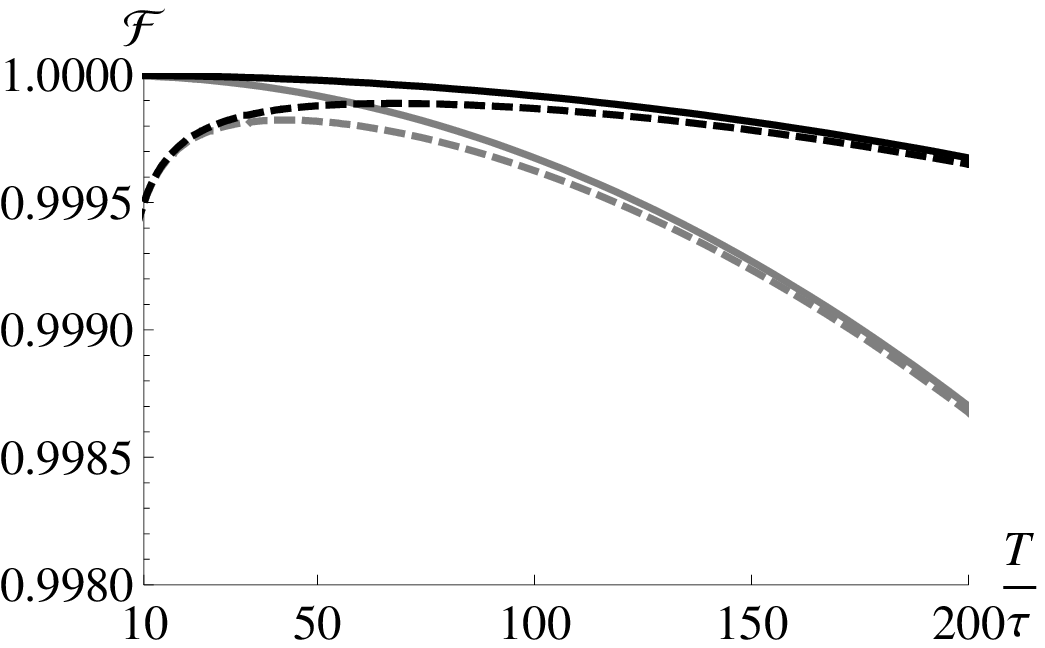}
                \caption{       \label{fig:twocoupled}
        Left: Entanglement as a function of $T/\tau$  for $\sigma=0.1$ and  
$\chi \tau=10^{-3}$ (grey dashed line);
        $\chi \tau=5~10^{-4}$ (black dashed line).
        Right: Fidelity as a function of $ T/\tau$ for $\sigma=0$ and $\chi  
\tau=10^{-3}$ (grey solid line),
        $\sigma=0$ and $\chi \tau=5~10^{-4}$ (black solid line),
$\sigma=0.1$ and $\chi \tau=10^{-3}$ (grey dashed line),
        $\sigma=0.1$ and $\chi \tau=5~10^{-4}$ (black dashed line).
        In all the plots the loop is as in Fig.~\ref{fig:thetaphiloop} and  
with the choosen $\theta_M$ and $\phi_M$ we have
        $\gamma = 0.75$.
        }

	\label{fig:coupled_qubs}
\end{figure*}

We shall consider as initial state the
maximally entangled state (\ref{maxent}). From
(\ref{cond1}), (\ref{cond2}), (\ref{maxent}) and (\ref{eq:dark_states_H_I}) it follows that
\begin{equation}
|\Psi \rangle=\frac{1}{\sqrt{2}}(|{\bar{D}}_1(0) 0\rangle + 
|{\bar{D}}_2(0)1\rangle)\,.
\label{initial}
\end{equation}
Note that the time evolution of $|{\bar{D}}_1(0)0\rangle$ is
driven by the unitary operator $U_0(T)$ in (\ref{eq:Hol_operator}), since
the interaction does not affect the state $|0\rangle$ of qubit B.  On
the {other hand}, $|{\bar{D}}_2(0)1\rangle$ evolves with the
perturbed unitary operator $U_\chi(T)$ in (\ref{eq:U_chi}). {Thus, to study the evolution of $\Psi$ it is useful to introduce the operator}
\begin{equation}
\mathcal U(T)=U_0(T)\otimes 
\left(
\begin{array}{ll}
 1&0\\0&0
\end{array}
\right)+U_{\chi}(T)\otimes\left(
\begin{array}{ll}
 0&0\\0&1
\end{array}
\right)\,.
\label{matu}
\end{equation}
Here, {as usual}, $U_0(T)$ and $U_{\chi}(T)$ act on the qubit A, while the other two matrices act on the qubit B
and  are expressed in the $|0\rangle,|1\rangle$ basis for B. {Then}
\begin{eqnarray}
\mathcal U(T)|{\bar{D}}_1(0)0\rangle\!\!&=&\!\!\cos\eta|0 0\rangle -\sin\eta|1 0\rangle\nonumber\\
\mathcal U(T)|{\bar{D}}_2(0)1\rangle\!\!&=&
\!\frac{e^{i \frac{\chi T \gamma}{2}}}{\mu }
  \!\!\left( 2 \eta  \sin \frac{\mu }{2}|01\rangle 
    + K_+\!|11\rangle\right)
\label{eq:TwoQubsNoNoise}
\end{eqnarray}

{Let us now evaluate (\ref{iar}) at $t=T$,} 
\begin{equation}\rho_A^r=
    \mbox{Tr}_B \left( P \otimes P \mathcal U(T)|\Psi\rangle\langle
      \Psi| \mathcal U^{*}(T)P\otimes P \right)\,,\end{equation}
{and thus, from (\ref{matu}),}
\begin{widetext}
\begin{equation}
 \rho_A^r=\frac{1}{2\mu}
\left(
\begin{array}{ll}
 \qquad\mu &\quad e^{-\frac{i \chi T \gamma }{2}} \left(2 \eta  \cos \eta  \sin \frac{\mu
   }{2}-K_- \sin \eta  \right) \\\\
 e^{\frac{i \chi T \gamma}{2}} \left(2 \eta  \cos \eta  \sin
   \frac{\mu }{2}-K_+ \sin \eta  \right) &\quad \qquad\mu
\end{array}
\right).
\label{densitychi}
\end{equation}
\end{widetext}

The calculation of $\mathcal{E}^r$ is now straightforward by using
(\ref{eq:entanglement}) and (\ref{densitychi}). Here, we quote the
lowest order expansions in the coupling $\chi$ 
\begin{equation}
\mathcal{E}^r = 1-\frac{(\gamma \chi T )^2\sin^4(\eta)}{8\eta^2}+O((\chi T)^3))\,.
\label{Eeta}
\end{equation}
Note that the presence of the qubits coupling affects the entanglement
inducing a quadratically decreasing behavior in  $\chi$. Only in the limit 
$\eta\to 0$ (no holonomic transformation) the
entanglement is still preserved, $\mathcal{E}^r = 1$, irrespectively on the
coupling. Indeed, in this case the time evolved state, starting from a maximally
entangled state (\ref{initial}) differs from the initial state by a phase factor
$|\Psi(T)\rangle=(|00\rangle + \exp(-i\chi T) |11\rangle)/\sqrt{2}$, which does not affect the entanglement.

The fidelity  (\ref{fidelity1}) can now be evaluated for $\mathcal{U}(T)$ given by   (\ref{matu}) (see also (\ref{eq:TwoQubsNoNoise})). {One finds}
\begin{eqnarray}
	\mathcal{F} &=& 1-
\left(7+8\eta^2+8\eta\sin(2\eta)-8\cos(2\eta)+\cos(4\eta)\right)\times \nonumber\\
&&\times \frac{ (\gamma\chi T)^2}{256 \eta^2}+O((\chi T)^3))\,.
\label{fidelitya}
\end{eqnarray}
{Note that}, differently from the entanglement, even in the absence of holonomic transformation ($\eta=0$)
the fidelity is affected by the coupling, {i.e.,}
$$\mathcal{F}\vert_{\eta=0}  = 1- \frac{(\gamma\chi T)^2}{8}+O((\chi T)^3))\,.$$

We shall now take into account the effect of the  parametric noise and consider the case of a perturbation to the logical NOT operation with $\eta=\pi/2$.
The perturbed unitary operator
$\widetilde{U}_\chi$  is obtained from (\ref{eq:U_chi}) by means of  the substitutions
\begin{eqnarray}
\eta
&\rightarrow& \widetilde\eta=\frac{\pi}{2} + \delta \eta\nonumber\\ 
\mu&\rightarrow&\widetilde{\mu}= \sqrt{(\pi+2 \delta \eta)^2+(\chi T
  \gamma) ^2}\nonumber\\
K_{\pm}&\rightarrow&\widetilde K_{\pm}=
\left(\tilde\mu \cos \frac{\tilde\mu }{2}\pm i \chi T \gamma \sin
  \frac{\widetilde\mu }{2}\right)\nonumber
\end{eqnarray}
and the evolution
operator which extends (\ref{matu}) is 
\begin{equation}
\widetilde{\mathcal U}(T)=\widetilde U(T)\otimes 
\left(
\begin{array}{ll}
 1&0\\0&0
\end{array}
\right)+\widetilde U_{\chi}(T)\otimes\left(
\begin{array}{ll}
 0&0\\0&1
\end{array}
\right)
\label{matu1}
\end{equation}
with $\widetilde U(T)$ given in Eq.~(\ref{eq:tildeU}), {whence} 
\begin{widetext}
\begin{equation}
 \rho_A^r=\frac{1}{2\widetilde\mu}
\left(
\begin{array}{ll}
\qquad\qquad\tilde\mu& 
-e^{-\frac{i \chi T \gamma}{2}} \left[\widetilde K_- \cos \delta \eta   +2 \tilde\eta  \sin \delta \eta \sin
   \frac{\tilde \mu }{2}\right]
\\
-e^{\frac{i \chi T \gamma }{2}} \left[ \widetilde K_+ \cos \delta \eta   + 2 \tilde\eta  \sin \delta \eta \sin
   \frac{\tilde\mu }{2}\right]
\qquad\qquad
   \tilde\mu
\end{array}
\right).
\label{rhofinale}
\end{equation}
\end{widetext}

Thus,  from (\ref{eq:entanglement}),  (\ref{fidelity1}), and (\ref{rhofinale}),  we finally obtain (at the  lowest orders in the errors) 
\begin{eqnarray}
\mathcal{E}^r &\approx& 1-\frac{\left(\pi -4 \delta\eta\right)  (\gamma\chi T)^2}{2 \pi ^3}
\label{eq:ent_two_qubs}\\
\mathcal{F} &\approx& 1-\frac{(\delta\eta)^2}{2 }-\frac{\left(8+\pi ^2\right) 
(\gamma\chi T)^2}{32
   \pi ^2}\,. \label{fidelity2}
\end{eqnarray}
It should be observed {that} $\mathcal{E}^r$ is jointly reduced by coupling $\chi$ and parametric noise $\delta\eta$ (yet  $\mathcal{E}^r =1 $ for $\chi =0$  and  $\delta\eta\neq 0 $, in agreement with the results of Sec.~\ref{sec:ent_fid_cal}), and that the parametric noise may increase the entanglement with respect to the bare case
with  $\eta=\pi/2$ (in accordance with the
behavior discussed in Eq.~(\ref{Eeta}), where $\mathcal{E}^r\to 1$ for
$\eta\to 0$).   Moreover, note that the fidelity $\mathcal{F}$ is influenced independently by the two sources of noise. Indeed, in addition to the decrease induced by  the
coupling, already {accounted by} (\ref{fidelitya}), the fidelity is also
depressed by the geometric perturbation $\delta \eta$ in agreement
with the results in Sec.~\ref{sec:ent_fid_cal}.

The 
model of parametric noise introduced in
Sec.~\ref{sec:evolution_with_noise} allows for 
a more quantitative analysis of  (\ref{eq:ent_two_qubs}) and (\ref{fidelity2}).
According to this model, 
$\delta \eta= \sigma \sqrt{\tau/T}$ (cf. Eq~(\ref{erroresueta})), with
$\sigma^2$ the variance and $\tau$ the time scale of the noise
fluctuations; the condition of small perturbations, $\delta \eta \ll
\pi/2$, implies $\sigma \sqrt{\tau/T} \ll \pi/2$. The  behaviors of entanglement and fidelity are represented in Fig.~\ref{fig:twocoupled}; more precisely: 
\begin{itemize}
\item Fig.~\ref{fig:twocoupled} (left) shows the entanglement
(\ref{eq:ent_two_qubs}) as a function of the final time $T$ for fixed
$\sigma$ and different values of $\chi\tau$.  The value of
$\gamma$, defined in (\ref{gamma}), depends on the loop chosen with
$0.66\le\gamma\le 1$. Here, $\gamma=0.75$.  The behavior
suggests a way to preserve entanglement: choosing not too long
evolution times the dominant source of error can be minimized.  In
particular an evolution with $T< 200\tau$ causes an error on the
entanglement smaller than $1.5~10^{-3}$.

\item Fig.~\ref{fig:twocoupled} (right) shows the fidelity. When $\sigma=0$, i.e. no parametric error,
the fidelity shows a quadratically decreasing behavior as a function
of $T/ \tau$ depending on the values of $\chi \tau$ (solid lines).  The presence of a parametric
noise changes qualitatively the behavior for small $T/\tau$.  In fact,
in this region independently of the $\chi$ coupling, the fidelity
drops because the geometric error dominates.  By increasing $T/\tau$
there is an intermediate region where the fidelity increases before
fast decreasing, when the $\chi$ error prevails.  Intermediate times
evolution are then the more efficient to preserve the fidelity. A good
choice is $50 \tau < T < 100 \tau$.  In this range of adiabatic times,
both entanglement and fidelity have errors of order $5~10^{-4}$.
\end{itemize}

To conclude, the constraints on the time scales of our model that
allow to construct holonomic gates which preserve both entanglement
and fidelity and are consistent with the adiabatic approximation are:
\begin{equation}
  \Omega^{-1} < 50 \tau < T < 100 \tau < \chi^{-1}
\end{equation}
with the additional request that $\sigma \ll \sqrt{50} \pi/2$.

\section{Conclusions}
\label{sec:conclusions}
In the present paper we studied a noisy two-qubit system in order
to understand if and how the holonomic operators preserve the
entanglement. We considered a model in which only one of the
two qubits undergoes a holonomic transformation.  Being the two
dimensional logical space embedded in an extended four dimensional
Hilbert space we introduced, as possible estimator, the reduced logical
entanglement which correponds to the fraction of entanglement in the
logical subspace.  We also calculated the fidelity and compared it with
the reduced logical entanglement. 

We considered two types of noise: a parametric noise that goes off
whenever the driving fields are turned off, and a coupling  noise due
to undesired interactions between the two qubits.  We have shown that the
holonomic operators are robust under parametric error.  In particular,
the entanglement is preserved under an holonomic transformations while
the fidelity is affected by such a noise but, due to geometric
cancellation effects, it can reach good values for long times. In the presence also of a coupling error,  we showed that
the entanglement is mainly influenced by this noise
and weakly by the geometric perturbation. Instead, the  fidelity shows
a different dependence  on the adiabatic time:
it is dominated by the parametric noise for not too large times and depends on the coupling error at larger times.
We demonstrated that the intermediate time evolutions are the
best choice.  Within a realistic range of physical parameters, we found
that for both entanglement and fidelity the error can be strongly
reduced.\\

{\bf Acknowledgments}\\

We thank E. De Vito and A. Toigo for {many fruitful} discussions. N.~Zangh\`\i\ was supported in part by INFN.


\begin{thebibliography}{99}
\bibitem{loopnote}
The dependence of the angle variable taken for the calculation is :
$\theta(t)=\{0, \theta _m+ \frac{4 t \left(\theta _M-\theta _m\right)}{T},\theta_M, -2 \theta _m+3 \theta _M-\frac{4 t \left(\theta _M-\theta
_m\right)}{T}, \theta_m\}$ and $\phi(t)=\{0,\left(\frac{4 t}{T}-1\right) \phi _{M } , \phi_M, 4 \left(1-\frac{t}{T}\right) \phi _{M }\}$
where the evolution is considered along the path in Fig.~\ref{fig:thetaphiloop}.
The corresponding ``speeds'' are $\dot{\theta}(t)=\{0, \frac{4  \left(\theta _M-\theta _m\right)}{T},0, -\frac{4  \left(\theta _M-
\theta_m\right)}{T}, 0 \}=\{0, v_{\theta},0, -v_{\theta}, 0 \}$ and $\dot{\phi}(t)=\{0,\frac{4 \phi_M}{T}   , 0, -\frac{4 \phi_M}{T}   \} =\{0,v_{\phi}   , 0, -v_{\phi}   \}$.
\end{thebibliography}
\end{document}